\documentstyle[aps,prl,twocolumn,psfig]{revtex}
\begin{document}
\draft
\title{Structured Mixed Phase {\it is} Favored in Neutron Stars}
\author{Michael B. Christiansen}
\address{Institute of Physics and Astronomy,
University of Aarhus,
DK-8000 \AA rhus C, Denmark}
\author{Norman K. Glendenning}
\address{Nuclear Science Division and
Institute for Nuclear \& Particle Astrophysics,
Lawrence Berkeley National Laboratory, MS 70A-319, Berkeley, California 
94720}
\date{\today}
\maketitle

\begin{abstract}
We give a general thermodynamical argument showing that in neutron stars,
the 
Coulomb structured mixed phase is always favored for any first order phase 
transition involving systems in equilibrium with baryon number and electric 
charge as the two independent components. This finding is likely to have 
important consequences for many neutron star properties, e.g., glitch 
phenomena, transport and superfluid properties, r-mode instabilities, and
the 
braking index.
\end{abstract}


A number of phase transitions may take place in neutron stars. Described in 
most detail is the first order nuclear liquid-gas transition in the inner
crust
of a neutron star, where nuclei with different shapes are in equilibrium
with 
a neutron gas \cite{pethick95,lorenz}. 
Above the nuclear saturation density, phase transitions involving, for
instance,
quark matter, pion, and kaon condensation have been suggested and described,
see e.g., Ref.\ \cite{heiselberg00} for an overview.

Before 1992, local charge neutrality was imposed on all the possible phase 
transitions above saturation density, leaving the systems with only one 
independent component, namely the baryon number. This treatment made any
first 
order transition resemble, e.g., the water-ice transition.  
However, in 1992 it was pointed out by Glendenning \cite{nkg92} that 
because neutron star matter has {\it two} independent components, namely the
baryon number and the electric charge, charge neutrality must be applied 
as a global and not a local condition.
Gibbs general conditions for thermodynamic equilibrium for both chemical 
potentials (usually taken to be the neutron and electron chemical
potentials) 
cannot be satisfied otherwise. As a consequence of this, the charge does not
vanish identically; rather, there results a nonvanishing charge density of 
opposite sign for each bulk phase \cite{glbook}.  

\begin{figure}[ht]
\vspace{-.2in}
\begin{center}
\leavevmode
\psfig{figure=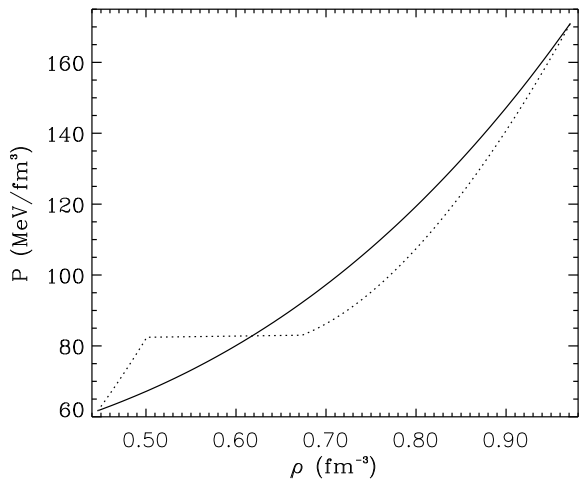,width=3.4in,height=2.76in}
\parbox[t]{5.5in} { \caption { \label{fig1} The pressure as a function of
the
baryon density in the mixed phase region of a two-component system. The
solid 
curve corresponds to the two-component system, whereas the dotted curve 
corresponds to the same system with the electric charge component frozen
out.
The actual numbers are for a system where the first order phase transition
is 
to a kaon condensate \protect\cite{glsb98}.}}
\end{center}
\end{figure}

A first order transition with multiple independent components behaves
qualitatively differently from a single component one as seen in Fig.\ 
\ref{fig1}. For a one-component system, the transition takes place at a 
constant pressure where the two phases, having different densities, for all 
proportions are in equilibrium. 
The equilibrium pressure is found through a Maxwell construction.
Systems with two components have an additional degree of freedom, which
means
the proportions of the two phases in equilibrium vary
as a function of the pressure. The phases still have different densities,
but
the densities are now also functions of the pressure
\cite{nkg92,callen,notea}.
Mixtures of two miscible liquids in equilibrium with their vapor phase are
well described examples from physical chemistry \cite{callen,hirsch}. 
In a neutron star this means that a two-component phase transition may have 
a considerable radial extent in contrast to a one-component transition, 
which takes place at a single radial point in the star, the point where the
pressure of the two phases are equal.
This region of two coexisting phases in equilibrium is usually referred to
as 
the mixed phase.

In \cite{nkg92} surface and Coulomb effects were neglected in the
description
of the mixed phase, although it was pointed out that their inclusion would
result in a Cou\-lomb lattice, analogous to the lattice resulting from the 
nuclear liquid-gas transition in the inner crust of a neutron star. 
Such a structured mixed phase above the saturation density was first studied
in \cite{pethick93} for the deconfinement transition. It was there concluded
that the mixed phase may not be energetically favored, compared to the 
locally charge neutral, one-component system, if the surface and  
Coulomb energies are `sufficiently large'. 

We show in the following that for any first order phase transition of a 
two-component system of nuclear matter in full thermodynamic equilibrium,
the 
energy (including Coulomb and surface energies) of this system is always 
smaller than the energy of the same system where one component is frozen out
\cite{noteb}. For a neutron star, the important implication is that for
matter  
in equilibrium there will always be a structured mixed phase if any first
order
phase transition takes place in its dense interior. 

Gibbs conditions for thermodynamical equilibrium between two 
phases (I and II) with $n$ independent components separated by an arbitrary 
boundary are \cite{gibbs,guggenheim}
\begin{eqnarray}
\label{equil} 
 T^I&=&T^{II}   \nonumber \\
 \mu^I_i&=&\mu^{II}_i \hspace*{1cm} i=1,2,...,n  \nonumber \\
 P^I(\mu_i^I,T^I)&=&P^{II}(\mu_i^{II},T^{II})+\sigma\left(\frac{1}{R_1}
 +\frac{1}{R_2}\right)+...,
\end{eqnarray}
where $T$ is the temperature, $\mu_i$ is the chemical potential of the
$i$'th 
independent component, $P$ is 
the pressure, $\sigma$ is the surface tension, and $R_1$ and $R_2$ are the 
principal curvature radii (higher order corrections like curvature effects
have
not been explicitly included in the pressure equilibrium condition). 
These conditions describe thermal, chemical, and mechanical equilibrium, 
respectively. In addition the energy (as well as the free energies) is at a 
minimum for a system in equilibrium.
All these conditions are direct consequences of the second law of 
thermodynamics.
In the case of a neutron star, Gibbs conditions have to be solved
consistently 
with the condition of charge neutrality, which as emphasized, cannot be 
imposed as a condition of local vanishing charge density else there are too 
many unknowns for the number of equations.

Cold nuclear matter in a single phase in equilibrium is of course
electrically 
neutral. $\beta$-equilibrium ensures that the sum of the proton and electron
chemical potentials equals the chemical potential of the neutron, 
$\mu_n=\mu_p+\mu_e$ (the neutrinos escape the system freely). 
This local charge neutrality requires the electron and proton densities to
be equal in bulk. 
Earlier, local charge neutrality was likewise imposed on both phases of the 
system during a phase transition. This assumption has the effect of freezing
out the electric charge component and leaving the baryon number as the
single 
independent component.
The Maxwell construction ensures pressure equilibrium and chemical
equilibrium 
between the baryons across the phase boundary for such a phase transition.
However, the electron chemical potential is generally discontinuous across
the 
phase boundary.

In the current picture of the bulk phases, the local charge neutrality of
both 
phases is relaxed 
to a global charge neutrality condition, which allows the system to freely 
explore the additional degree of freedom in a system with two independent 
components.  
Compliance with Gibbs phase equilibrium conditions for both chemical 
potentials naturally result in a nonvanishing charge density for each phase.
The normal nuclear matter phase will be positively charged, and thus 
more isospin symmetric; whereas the high density phase (e.g., quark matter)
will be negatively charged, and partly replace electrons (and muons) as
global 
neutralizing agents.
The bulk energy of the one-component system lies above the bulk energy of
the
mixed phase system if Coulomb and surface energies are neglected, since the
charge neutrality constraint is relaxed in the latter approach \cite{nkg92}.
\begin{figure}[ht]
\vspace{-.2in}
\begin{center}
\leavevmode
\psfig{figure=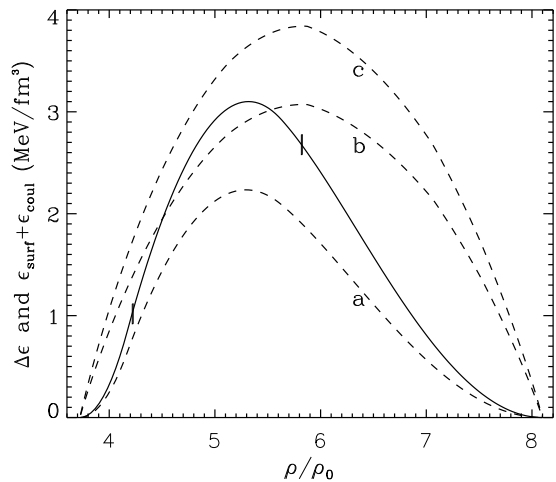,width=3.4in,height=2.76in}
\parbox[t]{5.5in} { \caption { \label{fig2} The solid curve is a schematic
plot of the difference in energy density between a Maxwell system and the
corresponding bulk mixed phase system as a function of the baryon
density in units of the saturation density. The two marks represent the
densities of the low and high density phase at the constant transition
pressure of the Maxwell system, between which the average baryon density
varies linearly according to $\rho=\chi\rho^I+(1-\chi)\rho^{II}$,
where $\chi$ is the volume fraction of phase $I$ \protect\cite{noted}.
The curves $a$, $b$, and $c$ are schematic plots of the sum of Coulomb and
surface energies in the mixed phase region corresponding to the three
possible scenarios described in the text.}}
\end{center}
\end{figure}

In the current picture of the structured mixed phase as presented in 
Ref.\ \cite{pethick93}, the mixed phase is energetically favored only
if the sum of Coulomb and surface energies in the mixed phase is smaller
than
the gain in bulk energy between the two systems. 
A schematic plot of this comparison is shown in Fig.\ \ref{fig2} (see also
the figure in \cite{pethick93} or Fig.\ 30 in \cite{heiselberg00}). 
There are three possible scenarios, corresponding to the three curves of 
Coulomb and surface energies: $a$, the structured mixed phase is favored in 
the whole bulk mixed phase region, $b$, it is favored in some regions of 
the bulk mixed phase, $c$, it is not favored anywhere. 
The properties of the equations of state including surface and screening 
effects for the two phases determine which scenario is valid for the phase 
transition in question.
Screening effects reduce the Coulomb energy and will play an increasingly 
important role with increasing 
size of the charged structures. Screening lengths for different components
have been estimated, the ranges are from about 5 fm to 13 fm. Structures
that
are smaller than the screening lengths will have almost uniform particle 
densities within each phase and a nonvanishing local charge density. 
On the other hand, the larger the structures are compared to the screening 
lengths, the more the system will resemble two electrically neutral bulk 
phases \cite{pethick93,heiselberg00}. 

The new picture we present is based on the thermodynamical equilibrium
conditions described in Eq.\ (\ref{equil}) and below. These conditions are 
direct consequences of the second law of thermodynamics and therefore have 
general validity.
The standard Maxwell construction can only ensure mechanical equilibrium 
and {\it one} chemical potential to be in equilibrium across a phase
boundary.
The other chemical potential will generally be discontinuous across 
the boundary. Thus, if one component is frozen out by imposing constraints
on 
a system with two independent components, this system cannot be in {\it full
thermodynamical equilibrium}. Therefore the (free) energy of such a 
system is not at a minimum, and there must be another system in
thermodynamical
equilibrium in which the energy is a global minimum. This statement is valid
over the whole range of the mixed phase region and not only at the pressure 
$P_{t}$, where the actual discontinuity in one chemical potential is.
To see this consider the system (I) in the neutron star with the global
energy 
minimum at $P_{t}$ 
and assume that the single phase (II) with the imposed local charge
neutrality is preferred at an infinitesimal smaller (or larger) pressure.
There
will be a phase boundary between these phases, but they will have different
chemical potentials and therefore will only be in mechanical equilibrium.
Thus
there must be a system different from (II) at this pressure which is in 
thermodynamical equilibrium with (I) and being at the global energy minimum.
This argument can be repeated until the whole mixed phase region has been
spanned.
Notice, that the above argument is valid irrespective of whether the central
pressure of the neutron star is larger than $P_t$ of the Maxwell system or
not.
Therefore, a two-component system with one component frozen out cannot be
at a global energy minimum since it is not in full thermodynamical
equilibrium.
The system with the global energy minimum, in the pressure region where the 
equations of state indicate a two-phase equilibrium is possible, must 
therefore be the mixed phase with the particular Coulomb lattice structure 
which has the lowest energy, since this is the only system which is in full 
compliance with Gibbs equilibrium conditions.

Screening and surface effects do not change the above conclusion. 
Phase boundaries are inevitable in two-phase systems in equilibrium, and 
likewise are screening effects an integral part of all charged systems in 
equilibrium, whether the systems consist of a single phase or two (or more) 
phases separated by a phase boundary.
In the latter case screening and surface effects across the
phase boundary become inseparable.
Charged systems larger than the typical screening length can reduce their 
Coulomb energy by reducing the local charge density.
But this does not imply that these large systems will generally
resemble that of a one-component system with the electric charge frozen out.
As long as the charge density does not vanish identically everywhere in the
systems, the Wigner-Seitz cells will survive \cite{notec}.
On this basis the scenarios $b$ and $c$ in Fig.\ \ref{fig2} can be excluded
for
any first order phase transition of such two-component systems in full 
thermodynamical equilibrium.

For a given phase transition a figure like Fig.\ \ref{fig2} is very useful 
though. 
If the model describing the structured mixed phase has an energy
corresponding
to scenario $b$ or $c$, we can immediately conclude that the model does not
give a correct description, since we already know the Maxwell-like, 
one-component approximation represents an excited state.
This is a powerful test of whether the approximations in the treatment of 
Coulomb and surface energies are good enough.   
An important implication is that a rough upper limit on the surface tension 
in a given model can be estimated from this comparison. However, one should 
be careful not to use (reasonable) guesses for the value of the surface 
tension to draw conclusions about the physical possibilities of the system
as it was done in \cite{pethick93}.
We note that in the description of the first order nuclear liquid-gas 
transition, the findings of Ref.\ \cite{lorenz} are that the structured 
mixed phase is indeed favored except close to the upper end of the 
mixed phase region. Thus, their model gives at least a physically sound 
description of the structures in most of the mixed phase region except close
to
its upper end where the details of the model, here becoming increasingly 
important, can be improved upon. 

There are situations where scenario $b$, if it had been favored, could have 
led to buoyancy instabilities analogous to a situation where mercury is
placed 
on top of liquid water. 
Assume for example that the structured mixed phase for the system shown in 
Fig.\ \ref{fig1} is favored only if the pressure is less than 65
MeV/fm$^{3}$ 
corresponding to baryon densities below 0.48 fm$^{-3}$.
Since hydrostatic equilibrium requires that the pressure increases
monotonically from the surface to the center of the neutron star, there will
be a region in the star where a denser structured mixed phase is located 
further from the center than a less dense Maxwell-like phase. This is a
gravitationally unstable situation commonly seen in some stars. 
A similar situation arises within the structured mixed phase, but here
the two phases in each Wigner-Seitz cell have a non-vanishing charge
density, 
and the Coulomb force which is much stronger than gravity stabilizes the 
cell and secure that it remains intact.

In specific studies of possible phase transitions, reasonable models for the
equations of state of the phases are crucial. 
For some choices of models (or parameter values within a model) it may
happen 
that it is not possible to assure full compliance with Gibbs equilibrium 
conditions, even if a Maxwell construction, indicating the transition is
first 
order, can be found. 
Examples relating to kaon condensation can be found in \cite{pons}.
Since the Maxwell construction represents an excited state in a 
two-component first order phase transition, the model must be unphysical 
since by construction it cannot describe the system in full thermodynamical 
equilibrium.
We note that the authors of Ref.\ \cite{pons} are aware of this.

There is a single case where the transition can be first order without there
being mixed phase. If it happens
that also the electron chemical potential is continuous at the constant
pressure transition of the Maxwell-like system, then there will be no mixed
phase (or more precise, the radial extent of the mixed phase region has
shrunk to nothing). But this is by no means related to screening effects or 
energy considerations between the Maxwell-like system and the mixed phase.
It would be a completely coincidental property of the two phases and the 
equations of state describing them, and therefore it has vanishing
probability.

We have, based on Gibbs equilibrium conditions, presented a new picture 
showing that in neutron stars, the structured mixed phase will always be 
energetically favored for arbitrary first order phase transitions involving 
systems in full thermodynamical equilibrium with baryon number and electric 
charge as the two independent components. 
Therefore, models that impose local charge neutrality on such two-component 
systems and thereby make the phase transition resemble that of one-component
systems cannot be physically sound. 
Additionally, we have shown a way of checking whether the applied models are
physically reasonable. 

This conclusion is likely to have important consequences for a broad range
of 
neutron star properties, e.g., glitch phenomena, transport and superfluid 
properties, r-mode instabilities, and the braking index, see 
\cite{heiselberg00} and references therein. 
Details concerning these phenomena depend of course on the specific
properties
of any first order phase transition taking place in the dense interior of 
neutron stars, especially the radial extent of the mixed phase region, the 
density at which it starts forming, and whether unsurmountable energy
barriers 
prevent the neutron star matter from actually achieving its equilibrium 
configuration.

M.B.C.\ thanks Jes Madsen for useful discussions.
M.B.C.\ was supported in part by the Carlsberg Foundation and Direkt\o r Ib 
Henriksens Fond. 
N. K. G.\ was supported by the Director, 
Office of Science, Office of High Energy and Nuclear Physics, 
Division of Nuclear Physics, and by the Office of Basic Energy
Sciences, Division of Nuclear Sciences, of the U.S. Department of Energy 
under Contract No. DE-AC03-76SF00098.

\begin{references}
\bibitem{notea} This continuous variation in composition with pressure has 
misled some people to describe such a two-component phase transition as an 
overall second order transition with local first order characteristics 
\cite{heiselberg00}, but it is simply the characteristics of a first order 
phase transition in a multi-component system.
\bibitem{noteb} This statement is stronger than the general physical 
principle stating that opening a degree of freedom to a system has the
effect 
of lowering the energy of the system, or leaving it unchanged, not raising
it;
because the system may already be in its optimum configuration with respect 
to this degree of freedom, which leaves the energy of the system unaltered.
\bibitem{notec} If the charge density vanishes identically in a Wigner-Seitz
cell, the two phases will separate due to gravity. But because of the 
infinitesimal difference of the gravitational field across the dimension of
the
cell, this will only happen if the local charge density is vanishing.
\bibitem{noted} In contrast to the figure in Ref.\ \cite{pethick93} we also
include the energy difference between the two systems in the region,
where the average baryon density of the Maxwell system varies linearly.
The energy density of the mixed phase is not constant
in this region, and therefore the curve between the two marks is not a
straight line.
Notice also that the surface energy corresponding to the phase boundary in
the Maxwell system has been ignored previously. We do not include it either,
but note that this energy, if included in the energy difference curve, will
add
an additional contribution in the region between the marks.
\end{references}
\end{document}